\documentclass[8pt,a4paper,twocolumn]{article}

\usepackage[switch]{lineno} 
\usepackage{multicol,lipsum}

\usepackage[]{cite}

\usepackage{subfiles}

\usepackage{abstract}
\usepackage[table, dvipsnames]{xcolor}%Added for table coloring
\definecolor{lightgray}{gray}{0.85}

\usepackage[top=1.8cm,left=1.7cm,right=1.7cm,bottom=1.75cm]{geometry}
\usepackage{relsize}
\usepackage{tabularx}
\usepackage{array}
\usepackage{wrapfig}
\usepackage{multirow}
\usepackage{booktabs}
\usepackage{caption}
\usepackage{graphicx}
\usepackage{xcolor}
\usepackage{comment}

\usepackage[affil-it]{authblk} 
\usepackage{etoolbox}

\usepackage{lmodern}
\usepackage{amsmath}
\usepackage{mathtools}

\makeatletter
\patchcmd{\@maketitle}{\LARGE \@title}{\fontsize{16}{19.2}\selectfont\@title}{}{}
\makeatother

\begin{document}

\title{\textbf{{Y-Z cut lithium niobate longitudinal piezoelectric resonant photoelastic modulator}}}

\author[1*]{Okan Atalar}
\author[1]{Steven Yee}
\author[2]{Amir H. Safavi-Naeini}
\author[1]{Amin Arbabian}

\affil[1]{\textit{Department of Electrical Engineering, Stanford University, Stanford, California 94305, USA}}
\affil[2]{\textit{Department of Applied Physics and Ginzton Laboratory, Stanford University, Stanford, California 94305, USA}}
\affil[*]{Corresponding author: okan@stanford.edu\vspace{-2em}}

\date{}
\twocolumn[
  \begin{@twocolumnfalse}
\maketitle

\thispagestyle{empty}

\begin{abstract}
The capability to modulate the intensity of an optical beam has scientific and practical significance. In this work, we demonstrate Y-Z cut lithium niobate acousto-optic modulators with record-high modulation efficiency, requiring only $1.5~\text{W}/\text{cm}^2$ for 100\% modulation at 7~MHz. These modulators use a simple fabrication process; coating the top and bottom surfaces of a thin lithium niobate wafer with transparent electrodes. The fundamental shear acoustic mode of the wafer is excited through the transparent electrodes by applying voltage with frequency corresponding to the resonant frequency of this mode, confining an acoustic standing wave to the electrode region. Polarization of light propagating through this region is modulated at the applied frequency. Polarization modulation is converted to intensity modulation by placing the modulator between polarizers. To showcase an important application space for this modulator, we integrate it with a standard image sensor and demonstrate 4 megapixel time-of-flight imaging.\vspace{2em}
\end{abstract}

\end{@twocolumnfalse}
]

\section{Introduction}
Intensity modulators are fundamental components in optics with broad application spaces. Resonant free-space intensity modulators are a subclass of intensity modulators, allowing modulation of the intensity of free-space beams at a single frequency. The active area, modulation frequency, insertion loss, acceptance angle and modulation efficiency are some of the key performance metrics for resonant free-space intensity modulators. Free-space resonant intensity modulators with centimeter square scale area, megahertz modulation frequency, high modulation efficiency, and low insertion loss are crucial components for applications including wide-field lock-in detection~\cite{fluorescence_nature_imaging,video_rate_imaging_science,electro_optic_lifetime,resonant_electro_optic_lifetime,optical_lock_in_GW} and phase shift time-of-flight (ToF) imaging relying on standard image sensors~\cite{miller2020large,optical_resonator,SWIR_ToF,ToF_atalar}. Higher modulation frequency translates to a higher ranging accuracy, and is therefore preferred for ToF imaging applications.

Many physical mechanisms exist that allow control over the intensity of free-space optical beams. The solid-state approaches can be broadly classified into three categories: electro-optic, liquid-crystal-based, and acousto-optic. Electro-optic approaches offer high speed (exceeding gigahertz frequencies), but face challenges in modulation efficiency or limited acceptance angle due to reliance on optical resonance to boost efficiency~\cite{reflection_control_metasurface,slm_steering_brongersma,gate_tunable_metasurface,graphene_optical_modulator,enhanced_eo_ln,eo_slm_organic,ge_si_quantum_well_eo,eo_polymer_metasurface,plasmonic_phase_mod_free_space}, whereas liquid crystals are limited in their modulation frequency to the kilohertz regime~\cite{LC_ultrafast,LC_1,LC_2,LC_3,ferroelectric_slm}. Acousto-optics is a promising pathway, especially for resonant modulators. Mechanically resonant devices can attain quality factors significantly greater than those of electronic resonators at room temperature, resulting in high modulation efficiencies for acousto-optic devices.

%owing to modulation capability into the gigahertz regime and the high quality factors attainable with acoustic resonators (compared to radio frequency resonators at room temperature), resulting in high modulation efficiencies.

Acousto-optic intensity modulators typically consist of a piezoelectric transducer bonded to a suitable crystal. The piezoelectric transducer is excited with a radio frequency (RF) signal, which launches acoustic waves into this crystal and modulates the incident light on the crystal. Existing acousto-optic modulators have historically been classified into Bragg cells~\cite{acousto_optics_fundamentals,anisotropic_bragg_diffraction} and photoelastic modulators~\cite{transverse_photoelastic_mod,new_design_PEM}. Bragg cells rely on the diffraction of light by the acoustic wave. This is a phase-matched interaction, resulting in the modulation of light only for a narrow range of angles of incidence. Typical photoelastic modulators are resonant devices, where the input aperture determines the resonant frequency of the device. Centimeter square scale input apertures result in fundamental resonant frequencies in the kilohertz regime, thus sub-millimeter apertures are required to reach megahertz frequencies.

We have recently demonstrated a new type of acousto-optic modulator that we refer to as a longitudinal piezoelectric resonant photoelastic modulator~\cite{longitudinal_piezo_photo_mod}. This class of modulators offer a simple fabrication process: coating the top and bottom surfaces of a suitable piezoelectric material with transparent surface electrodes. These modulators have allowed centimeter square areas, low insertion loss, operation at megahertz frequencies, and significantly higher modulation efficiency compared to state of the art. In contrast with typical photoelastic modulators, our longitudinal resonator's resonant frequency is determined by its thickness, which lies along the optical axis. This arrangement decouples the input aperture from the resonant frequency, enabling designs with simultaneously large input apertures and high resonant frequencies. This advance has allowed watt level power consumption modulation of light at megahertz frequencies with a centimeter square aperture.

Operating at high modulation frequencies while requiring low power to operate the modulator is critical for many applications, including wide-field lock-in detection and ToF imaging relying on standard image sensors. The ranging accuracy of ToF imaging with standard image sensors is directly proportional to the modulation frequency, and lower drive power (i.e., higher modulation efficiency) greatly expands the application space (due to power constraints for many applications). Previous demonstrations of free-space intensity modulators are unable to reach high modulation frequencies and high modulation efficiencies, including our recent illustration of a modulator applied to ToF imaging~\cite{longitudinal_piezo_photo_mod}, which is limited by the unavoidable electrical losses in the modulator’s transparent electrodes. Transparent electrodes are essential for the propagation of light through our device, but have high electrical resistivity. The mechanical resistance of our device is inversely proportional to the product of the resonant frequency squared, quality factor ($Q$), and the active area. Simultaneous high operating frequency and large $Q$ (to achieve high modulation efficiency) result in a mechanical resistance that is orders of magnitude smaller than the electrode resistance. This makes it impossible to increase the resonant frequency and the $Q$ at the same time because the electrode resistance dominates the mechanical resistance.

The key feature missing in existing free-space intensity modulators is simultaneous high modulation efficiency and high frequency of operation. In this work, we introduce Y-Z cut lithium niobate (LN) longitudinal piezoelectric resonant photoelastic modulators to address this problem. These modulators allow reaching high modulation frequencies and high modulation efficiencies. Using this modulator, we demonstrate record-high modulation efficiency of $1.5~\text{W}/\text{cm}^2$ for 100\% modulation at 7~MHz for 532~nm wavelength light. This is an improvement in modulation efficiency by more than a factor of 17 compared to existing free-space intensity modulators and our previous demonstration that operate at megahertz frequencies. Using this modulator, we demonstrate 4 megapixel ToF imaging on diffuse reflectors using a standard CMOS image sensor. Our demonstration sets a record in modulation efficiency and opens a promising path forward for the design of low-power, large area resonant intensity modulators operating in the megahertz frequency regime that can find use for a plethora of applications, especially for high spatial resolution ToF imaging.

\section{Modulation Principle}
We provide an overview of the modulation principle for the modulator in this section. The modulator consists of a lithium niobate wafer coated on top and bottom surfaces with transparent electrodes. To operate the modulator, RF signal with frequency corresponding to the fundamental resonant frequency ($f_r$) of the modulator is applied to the surface electrodes to excite the fundamental shear resonance mode in the electrode region. Light propagating through the electrode region of the modulator is polarization modulated at the applied RF frequency, which can be expressed in terms of the static phase $\phi_s$ and the dynamic phase change $\phi_D$ (between the excited ordinary and extraordinary waves in the wafer). These phase terms are expressed in equation~\eqref{Eq.R1} and equation~\eqref{Eq.R2}, where $L$ is the modulator thickness, $r$ is the radius of the active region of the modulator, $S'_{yz}$ the excited strain amplitude in the rotated coordinate frame (see Fig. 1b), $n_o$ is the ordinary refractive index of LN, $n'_y$ the refractive index experienced by the extraordinary wave, $\lambda$ the wavelength of light, $p'_{14}$ and $p'_{24}$ the rotated photoelastic tensor components, and the volume integral carried out over the volume $V$ corresponding to the electrode region of the wafer (see Supplementary section 4 for more details).

\begin{gather}
\phi_s = \frac{2 \pi L(n_o - n_y')}{\lambda} \label{Eq.R1}
\end{gather}

\begin{gather}
\phi_D = \frac{2 \pi L}{\lambda} \Big(n_o^3 p'_{14} - {n'_y}^3 p'_{24}\Big)\sqrt{\frac{\int_V S'^2_{yz}dV}{\pi r^2 L}} \label{Eq.R2}
\end{gather}

Intensity modulation is achieved by placing the modulator between polarizers. The incoming unpolarized laser beam having perpendicular incidence to the wafer surface with intensity $I_0$ is intensity modulated. The intensity of the modulated laser beam is represented as $I(t)$, where $t$ stands for time, and is expressed in equation~\eqref{Eq.R3}, where $J_0$ and $J_1$ are the zeroth and first order Bessel function of the first kind, respectively; HOH stands for the higher order harmonics. 

\begin{gather}
I(t) = \frac{I_0}{2}\Big(\frac{1}{2} + \frac{1}{2}\big[\text{cos}(\phi_s)(J_0(\phi_D) - \nonumber \\ 2\text{sin}(\phi_s)J_{1}(\phi_D)\text{cos}(2 \pi f_r t) + \text{HOH}\big]\Big) \label{Eq.R3}
\end{gather}

\section{Modulator Design and Fabrication}
\subsection{Design}
We begin the design of the modulator by choosing the active area for the modulator (radius $r$) and the desired intensity modulation frequency ($f_r$). The two parameters that need to be determined are the modulator thickness ($L$) and the cut angle from Z to Y axis of lithium niobate, which we denote as $\beta$. Due to large aspect ratio (area to thickness) and anisotropic acoustic propagation, a pure mode cannot be excited with circular electrodes. This makes it difficult to accurately calculate the resonant frequency with dominant $S_{yz}$. However, we observe that the stiffness coefficient coupling to $S_{yz}$ does not vary strongly as a function of $\beta$ (see Supplementary Fig. S4). Therefore, the phase velocity for $S_{yz}$ can be approximated through $\sqrt{\frac{c_{44}}{\rho}}$, where $c_{44}$ is the stiffness coefficient, and $\rho$ the density of LN. A thickness of $L_i = \frac{\sqrt{\frac{c_{44}}{\rho}}}{2f_r}$ is initially chosen. Using this approximate thickness, we next determine the cut angle for the wafer. 

The cut angle of Y-Z LN is critical for controlling the mechanical resistance of our resonator, which should be significantly greater than the electrode resistance in order to operate efficiently. Electrode resistance is a challenge in our device due to the use of transparent electrodes, which are essential to the propagation of light through the device but are high resistivity compared to non-transparent conductors.

%The use of transparent electrodes to allow light to propagate through the modulator and interact with the acoustic standing wave in the wafer leads to the electrode resistance being problematic if the effective mechanical resistance is comparable to the electrode resistance. 

%To overcome this problem, the effective mechanical resistance for the piezoelectric resonator derived using the Butterworth Van-Dyke (BVD) model (see Supplementary section 3) can be controlled by adjusting $\beta$. 

To overcome this problem, we derive a BVD model of the piezoelectric resonator using rotated material parameters. We observe that the BVD model mechanical resistance  can be effectively controlled by adjusting the cut angle $\beta$ (see Supplementary section 3).

\begin{figure*}[t!]
\centering
\includegraphics[width=1\textwidth]{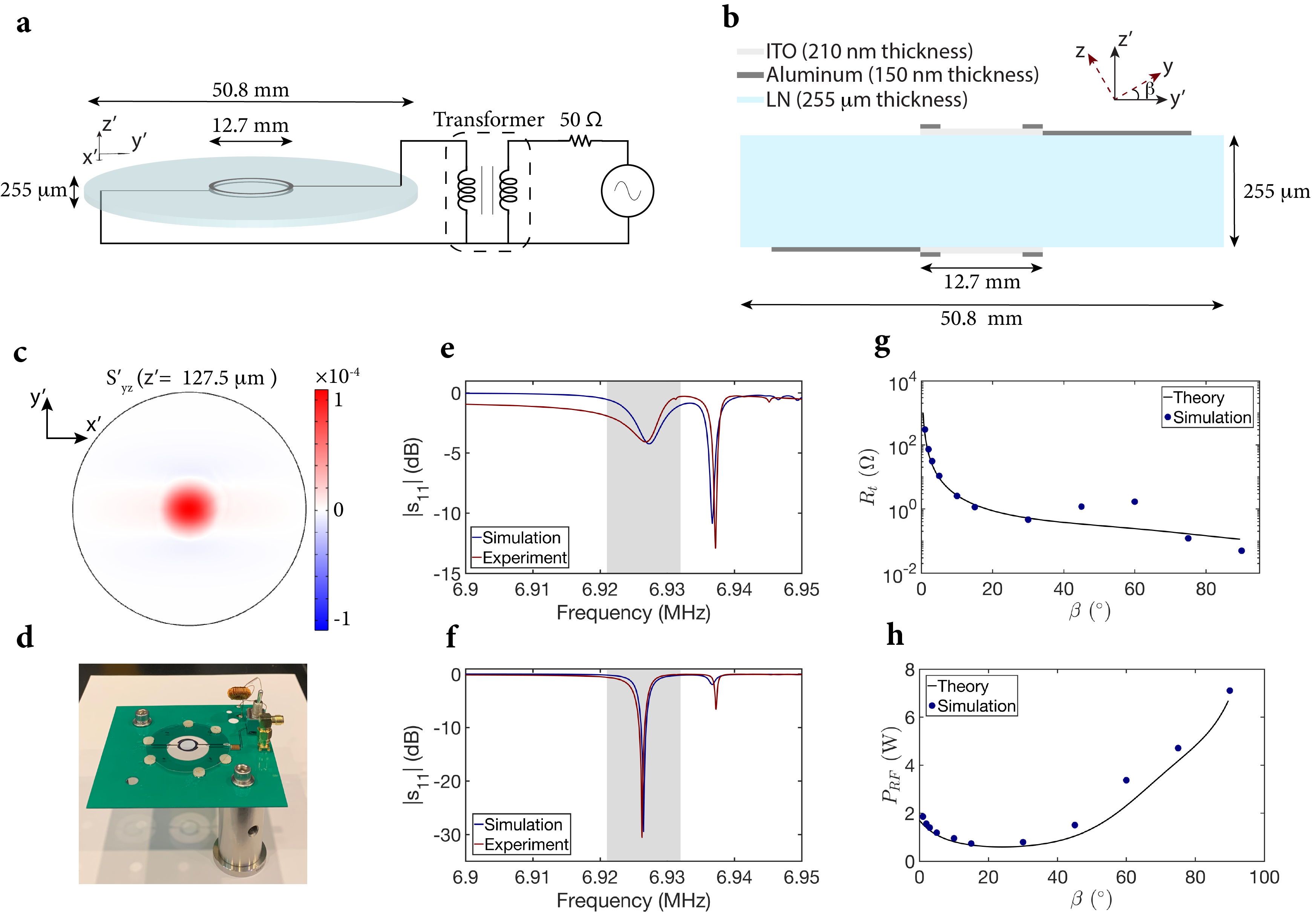}
\caption{Modulator design and electromechanical characterization. (a) A lithium niobate wafer having a diameter of 50.8~mm and a thickness of 255~$\mu $m is coated on top and bottom surfaces with transparent electrodes of diameter 12.7~mm. The top and bottom electrodes are connected to an RF power supply via an impedance matching transformer. (b) Side view of the lithium niobate wafer shown in (a). ITO is used for the transparent top and bottom electrodes, and aluminum strips on top and bottom surfaces are used to carry the RF power from the source to the ITO coated region of the wafer. The material coordinate system is shown with (x,y,z), and the rotated coordinate system with primed notation ($x'$,$y'$,$z'$), where $\beta$ is the rotation angle along the yz axis. (c) The simulated distribution of the dominant strain component  amplitude ($S'_{yz}$) in the primed coordinate frame when the wafer is excited at 6.926~MHz with 2Vpp applied to the surface electrodes is shown for the center of the wafer. $\beta = 5^{\circ}$ and a quality factor of $6 \times 10^3$ is used for the simulation to match experimental results. (d) The fabricated modulator mounted and wirebonded to a PCB is shown. (e) Simulated (blue) and experimental (red) of the device scattering parameter $|s_{11}|$ is shown  when no impedance matching  is used. The desired mode is highlighted in gray. (f) Simulated (blue) and experimental (red) of the device scattering parameter $|s_{11}|$ is shown when a transformer with 37 turns in the primary, and 18 turns in the secondary is used. The desired mode is highlighted in gray. (g) Simulated equivalent series resistance is plotted alongside model results as a function of $\beta$ with $Q = 6 \times 10^3$. (h) Model and simulation results for the required RF power to drive the modulator at 100\% optical intensity modulation are plotted against the cut angle $\beta$ (where $\phi_D = 1.2$) with $Q = 6 \times 10^3$.}
\label{fig:s1}
\end{figure*}

The sum of the mechanical resistance ($R_m$) and the electrode resistance ($R_s)$ is approximated in equation~\eqref{Eq.R4} in terms of the modulator parameters ($R_t = R_m + R_s$), where $e'_{34}$ is the rotated piezoelectric stress constant, $c'_{44}$ the rotated stiffness coefficient, and the phase velocity for the acoustic wave is $v' = 2 L f_r$. The primed notation indicates that we are in the rotated coordinate frame (see Fig.~\ref{fig:s1}). We note that the phase velocity $v'$, and therefore the resonant frequency, is relatively stable for small variations of $\beta$ around 0 degrees (Z-cut), whereas $R_t$ varies quite considerably (therefore $R_m$ varies considerably, since $R_s$ is not affected by $\beta$). Consequently, controlling $\beta$ serves as a powerful control knob over $R_m$ without significantly altering the other properties of the modulator. 

%To operate the device efficiently, $R_m$ should be larger than $R_s$ to attain as high $Q$ as possible.

\begin{gather}
R_t \approx \frac{c'_{44}v'}{4f_r^2 \pi^2 r^2 e'^2_{34}Q} \label{Eq.R4}
\end{gather}

%The only parameter that is unknown is the modulator quality factor $Q$. $Q$ depends on many parameters, but is highly dependent on the electrode thickness and electrode resistance. Knowing the relevant parameters, the cut angle can be chosen to adjust $R_m$. This value should be chosen to ensure that the mechanical resistance is not significantly smaller than the electrode resistance. Ideally, the mechanical resistance should be larger than the electrode resistance to limit power loss in the form of heat in the electrodes. The equivalent mechanical resistance can then be matched to the source via an impedance matching transformer.

%ith the chosen cut, the device is simulated in COMSOL with the initial thickness $L_i$. The resonant frequency ($f_{i}$) is identified corresponding to $S_{yz}$ by inspecting the admittance as a function of frequency in simulation. The final thickness can now be determined using scale invariance as: $L = L_i\frac{f_{i}}{f_r}$ to reach the desired modulation frequency of $f_r$.

For our design, we choose an active area radius of $r = 5~\text{mm}$ with a desired operation frequency of approximately 7~MHz, which is the highest frequency we can achieve for the $S'_{yz}$ mode based on wafer availability ($250~\mu\text{m}$ is the thinnest wafer we could obtain with customized cut angle). We chose $r = 5~\text{mm}$, since an important application space for these modulators is ToF, with the intention of placing the modulator in front of the image sensor pixel array or the lens of a camera. Standard image sensors offering megapixel resolution have centimeter square scale areas, necessitating similar dimensions for the active area of the optical modulator to make use of all the pixels while also having a large acceptance angle. Similarly, standard image sensors have lenses with centimeter square scale areas, requiring similar dimensions for the active area of the modulator to have high light collection efficiency.

In addition to $R_t$, another critical metric of our modulator is the RF power required ($P_{RF}$) to operate the device at some root mean square (rms) $S'_{yz}$, expressed in equation~\eqref{Eq.R5} (see Supplementary section 4 for derivation). To achieve high modulation efficiency, $P_{RF}$ should be made as small as possible when operating the device at 100\% optical intensity modulation (rms $S'_{yz}$ to make $\phi_D = 1.2$). Therefore, in the design of the modulator, both $R_m$ and $P_{RF}$ should be taken into consideration when choosing $\beta$.  

\begin{gather}
P_{RF} \approx \frac{4 \pi f_r c'_{44} \int_V S'^2_{yz}dV}{Q}
\label{Eq.R5}
\end{gather}

Our initial estimate of the quality factor is $Q = 1 \times 10^3$ for this modulator (based on our previous device). To achieve $R_t \approx 60~\Omega$ while also attaining high modulation efficiency (i.e., small $P_{RF}$), we choose $\beta = 5^\circ$. We want to keep our mechanical resistance larger than the electrode resistance to operate the device efficiently (i.e., have high $Q$). 

Choosing the right electrode thickness is important for achieving high modulation efficiency for these modulators. Thicker electrodes have lower resistance, and therefore allow smaller loss of RF power to the electrodes in the form as heat~\cite{60ghz_resonators,piezoelectric_AlN}. However, thicker electrodes also lead to larger acoustic attenuation and therefore lower $Q$, since acoustic waves are attenuated heavily in metals and alloys.  

With the chosen cut, the device is simulated using finite element modeling software (COMSOL Multiphysics)~\cite{COMSOL5} with the initial thickness $L_i$. The resonant frequency ($f_{i}$) is identified corresponding to $S'_{yz}$ by inspecting the admittance as a function of frequency in simulation. The final thickness can now be determined using scale invariance as: $L = L_i\frac{f_{i}}{f_r}$ to reach the desired modulation frequency of $f_r$.

The dominant strain component in our desired resonant mode is confined to the electrode region according to COMSOL simulation, as seen in Fig.~\ref{fig:s1}c . The other strain components in the wafer are shown in Supplementary Fig. S1. These other strain components have a significantly smaller power than the dominant $S'_{yz}$ component. In order to simplify the working principle in the preceding design equations, we assumed that only $S'_{yz}$ is excited in the wafer.

%while simultaneously being matched to the $50~\Omega$ impedance of the source.

It should be noted that cut angle can accommodate $Q$ values within some range through an impedance matching network. The reason why we do not choose $\beta$ to be sufficiently small (so that $R_m$ is always large) is to improve our modulation efficiency, and therefore reduce the required RF power to drive the device (see Fig.~\ref{fig:s1}h). 

\subsection{Fabrication}
We coat the top and bottom surfaces of a double-side polished Y85 ($\beta = 5^\circ$) lithium niobate wafer having a thickness of 255~$\mu $m and 5.08~cm diameter with 210~nm thick indium tin oxide (ITO) electrodes (to limit $R_s$ to several ohms) of radius 6.35~mm. The ITO was deposited in a load locked chamber using sputter coating. The initial sheet resistance of the ITO electrodes was $43~\Omega/\text{sq}$, and this was improved by heating at room pressure. Since Y85 LN is highly pyroelectric (charge accumulation along the top and bottom surfaces when heated), the heating of the wafer was performed while shorting the top and bottom surfaces of the wafer with an aluminum foil. The wafer covered with the aluminum foil shorting the top and bottom surfaces was heated on a hot plate; the temperature was raised by $5^\circ \text{C}$ every two minutes starting from $20^\circ \text{C}$ until $230^\circ \text{C}$ was reached. The temperature was then brought back to $20^\circ \text{C}$ by reducing the temperature by $5^\circ \text{C}$ every two minutes until $20^\circ \text{C}$ was reached. The final sheet resistance of the ITO electrodes was $22~\Omega/\text{sq}$.

%We chose to deposit 210~nm thick ITO for the transparent electrodes to make a reasonable trade-off between the internal quality factor of the wafer (without electrode resistance) and the electrode resistance. 

%We chose 210~nm thick ITO, since our previous device used a similar thickness and achieved a Q of $1 \times 10^3$. This electrode thickness will result in several ohms of resistance, and choosing $\beta = 5^\circ$ would result in $R_m$ of around $60~\Omega$, almost an order of magnitude larger than the expected electrode resistance.

To apply RF signal to the electrodes, for both the top and bottom surfaces of the wafer, we evaporate a 1~mm wide aluminum region around the edge of the ITO circular electrodes as well as a microstrip extension with width 1~mm to the edge of the wafer for wirebonding. 150~nm thick aluminum was coated on the wafer through evaporation in a load locked chamber. The aluminum allows the wirebonding and mechanical supports to be sufficiently far away from the acoustic mode which has a negligible strain profile outside the center 6.35~mm radius of the device, and therefore allows us to limit anchor losses. Additionally, the aluminum ring makes the active aperture of the modulator approximately 5~mm, since the aluminum at the edge of the circular ITO region is opaque.

\section{Modulator Characterization}

\begin{figure*}[t!]
\centering
\includegraphics[width=1\textwidth]{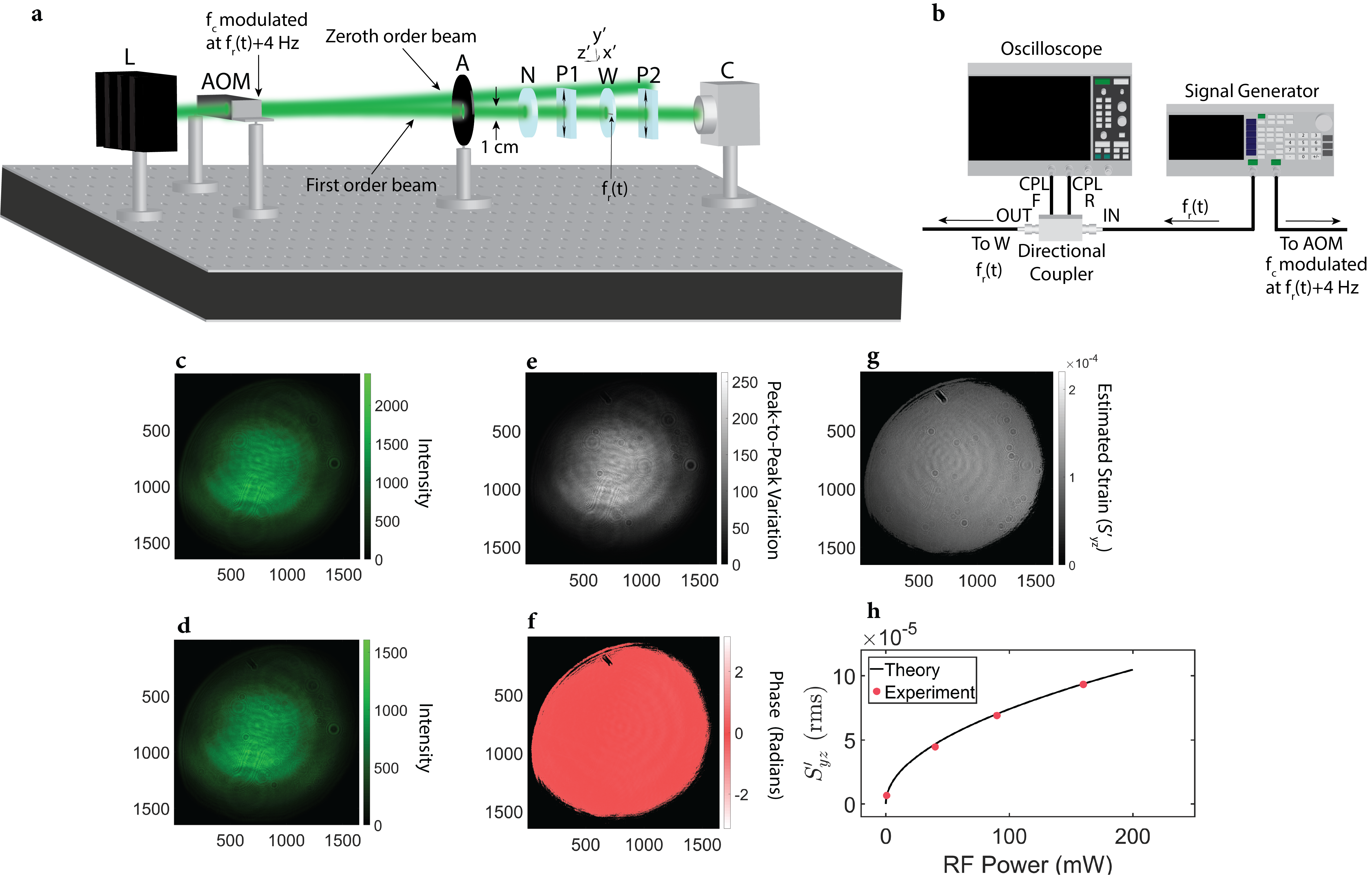}
\caption{(a) Schematic of the characterization setup. The setup includes a laser (L) with a wavelength of 532~nm that is intensity modulated at $f_r(t) + 4~\text{Hz}$ via an AOM (by modulating the carrier frequency $f_c = 80~\text{MHz}$),
aperture (A) with a diameter of 1~cm, neutral density filter (N), two polarizers (P1) and (P2) with transmission axis 
$\mathbf{\hat{t} = (\hat{a}'_x + \hat{a}'_y)}/\sqrt{2}$, the modulator
(W), and a standard CMOS camera (C). The modulator (W) is excited with an RF source of frequency $f_r(t)$, and the laser beam passes through the center of the wafer that is coated with ITO. The camera detects the intensity
modulated laser beam. (b) Equipment used for tracking the resonant frequency of the modulator is shown. The setup includes a signal generator driving the AOM using its first channel with center frequency $f_c$ amplitude modulated at $f_r(t) + 4~\text{Hz}$, and driving the LN modulator via a directional coupler using its second channel sending RF power at frequency $f_r(t)$ to the modulator. The oscilloscope detects both the input RF power to the modulator and reflected RF power from the modulator via a directional coupler. The oscilloscope and signal generator are interfaced to a computer running a MATLAB script that adjusts $f_r(t)$ with a negative feedback loop based on the phase difference between reflected and input waveforms. (c) Time-averaged intensity profile of the laser beam detected by the camera per pixel is shown when 160~mW of RF power at $f_r(t)$ is applied to the modulator and the second polarizer P2 is removed. (d) Time-averaged intensity profile of the laser beam detected by the camera per pixel is shown when 160~mW of RF power at $f_r(t)$ is applied to the modulator and the second polarizer P2 is present. (e) The peak-to-peak variation at 4~Hz of the laser beam is shown per
pixel when 160~mW of RF power at $f_r(t)$ is applied to the modulator and the second polarizer P2 is present. (f) The phase of intensity modulation at 4~Hz of the laser beam is shown per
pixel when 160~mW of RF power at $f_r(t)$ is applied to the modulator and the second polarizer P2 is present. (g) Estimated strain amplitude ($S'_{yz}$) per pixel is shown when 160~mW of RF power at $f_r(t)$ is applied to the modulator and the second polarizer P2 is present. (h) Estimated (Experiment) and theoretical (Theory) root mean square $S'_{yz}$ averaged over the modulator region (diameter of 1~cm) is shown for varying levels of RF excitation power.}
\label{fig:s2}
\end{figure*}

\subsection{Electromechanical Characterization}
We first characterize the fabricated modulator electromechanically. To perform this characterization, we mount the double sided electrode coated wafer on a printed circuit board (PCB), and wirebond the microstrip region on the top surface to the signal line of the PCB and the microstrip region on the bottom surface to the ground line of the PCB. We measure the $s_{11}$ reflection scattering parameter with respect to 50~$\Omega$ using a vector network analyzer (VNA) with excitation power of 0~dBm and bandwidth of 50~Hz. From this measurement, the $Q$ and $L$ (by comparing with COMSOL) of the modulator is extracted. Using this $Q$ and $L$, the device is simulated again in COMSOL. The overlaid simulation and experiment $|s_{11}|$ is shown in Fig.~\ref{fig:s1}e, showing a good match. The extracted $Q$ is $6 \times 10^3$ (higher than we  expected during the design phase of the modulator), with extracted $R_t = 11.5~\Omega$. To impedance match the resonator to the input source having a characteristic impedance of $50~\Omega$, we use a transformer. We connect a toroidal core transformer with 38 turns in the primary, and 17 turns on the secondary. We measure the $s_{11}$ of the modulator again using a VNA with the impedance matching transformer. Fig.~\ref{fig:s1}f shows the overlaid simulation and experiment $|s_{11}|$. The impedance matching transformer allows us to couple more than 99.9\% of the RF power from the source to the modulator (including electrode resistance $R_s$). The fabricated device mounted on a PCB with the impedance matching transformer is shown in Fig.~\ref{fig:s1}d.

\subsection{Optical Characterization}
To estimate the spatial strain profile in the wafer and to measure the modulation efficiency we perform optical measurements.

\begin{figure*}[t!]
\centering
\includegraphics[width=1\textwidth]{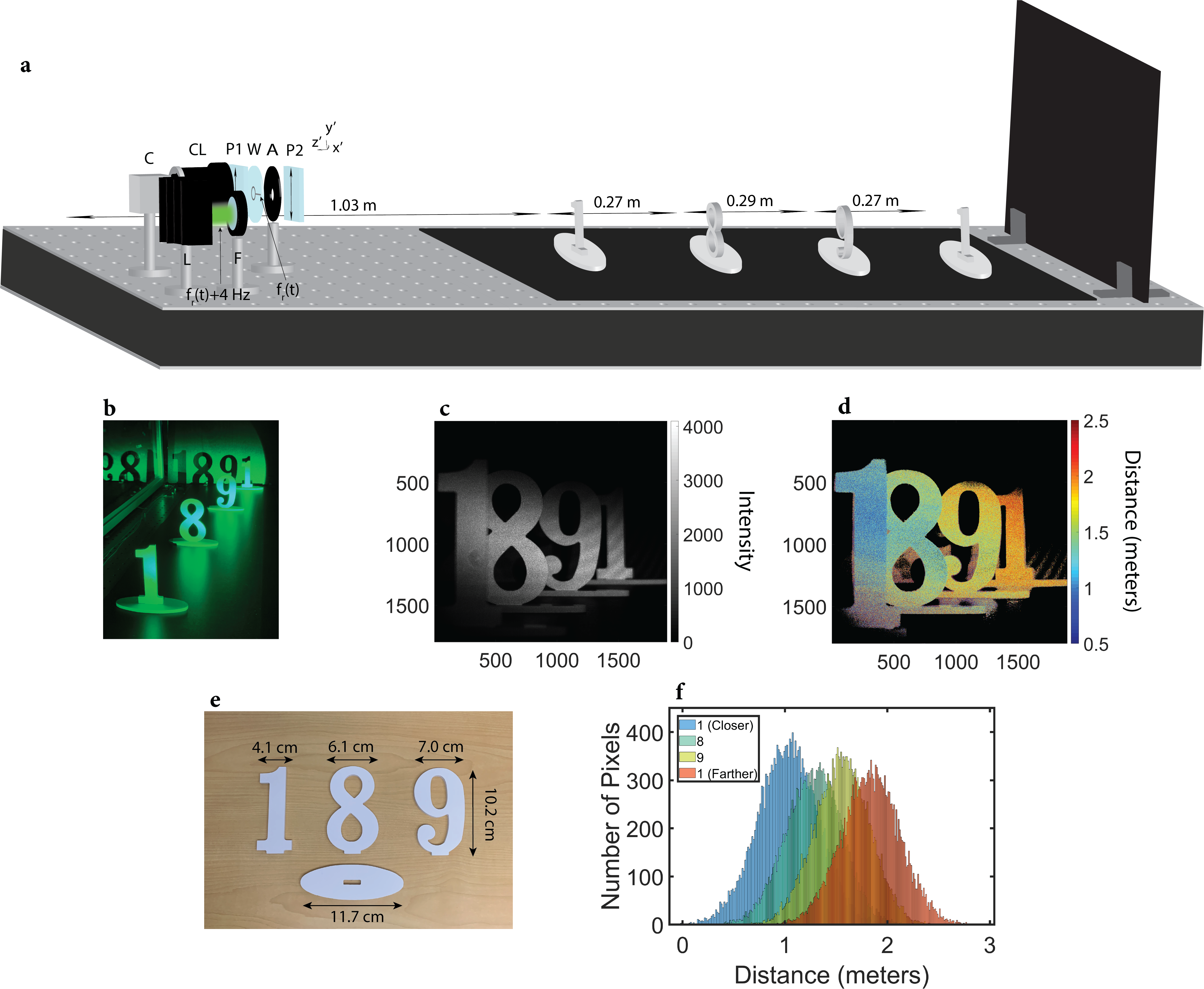}
\caption{(a) Schematic of the time-of-flight imaging setup. The setup includes a laser (L) with a wavelength of 520~nm that is intensity modulated at $f_r(t) + 4~\text{Hz}$,
aperture (A) with a diameter of 1~cm, two polarizers (P1) and (P2) with transmission axis $\mathbf{\hat{t} = (\hat{a}'_x + \hat{a}'_y)}/\sqrt{2}$, the modulator
(W), camera lens (CL), and a standard CMOS camera (C). The modulator (W) is excited with an RF power of 160~mW at frequency $f_r(t)$, and the laser beam is directed to targets placed away from the laser and the camera. A beam shaping lens (F) is placed in front of the laser to adjust the illumination area of the laser. (b) The targets illuminated with the laser are shown. (c) Time-averaged intensity detected by the camera per pixel is shown. (d) Reconstructed depth map per pixel by the camera.
Depth reconstruction is performed by converting the phase of the beat tone at 4~Hz to distance using equation~\eqref{Eq.R6}. Pixels that receive very few photons are displayed in black. (e) Dimensions of the targets used for the imaging experiment. (f) Distance distribution of the pixels corresponding to the different targets are shown. Approximately 20,000 pixels are used for each target to construct the histogram. The mean distance estimated by averaging the pixels corresponding to the targets are 1.03~m, 1.28~m, 1.56~m, and 1.84~m, respectively. The standard deviation of distance per pixel for the targets are 29.7~cm, 27.4~cm, 27.0~cm, and 28.8~cm, respectively.}
\label{fig:s3}
\end{figure*}

To reconstruct the dominant shear strain profile in the wafer ($S'_{yz}$), and to determine the modulation efficiency (RF power required to reach a certain volume rms strain level), we measure the intensity modulation imparted on a laser beam that propagates through the electrode region of the modulator with perpendicular incidence. Our setup is similar to that used in the characterization of our previous device~\cite{longitudinal_piezo_photo_mod}. A laser beam of wavelength 532~nm and with output power 10~mW is intensity modulated using an acousto-optic modulator (AOM) at $f_r(t) + 4$~Hz.  We take the first order diffracted beam as the modulated output. Modulation is achieved by amplitude modulation of an 80~MHz carrier RF waveform that drives the AOM (see Fig.~\ref{fig:s2}a). We measure a depth of modulation of 48\% for the intensity modulated laser beam using a fast photodetector. This intensity modulated beam is adjusted using a 1~cm diameter aperture to match the laser beam area to the active area of the LN modulator. The intensity modulated laser beam passes through the active region of the LN modulator and is captured via heterodyne detection with a standard CMOS image sensor having 4 megapixel resolution. 

%DoM for 532 nm laser estimated using 12_14_2021_AOM_Depth_of_Modulation. Compare peak-to-peak variation to DC value (using FFT) for 7 MHz case

Essentially, the spatial $S'_{yz}$ strain profile in the wafer is mapped to the spatial intensity profile of the optical beam at the output of polarizer P2, as shown in equation~\eqref{Eq.R3}. The use of a standard CMOS image sensor allows megapixel spatial resolution for reconstruction, but the sensor has an insufficient frame rate to capture $f_r$ directly. This is the reason behind using heterodyne detection; $S'_{yz}$ per pixel is extracted from the 4~Hz beat tone. The optical characterization setup is shown in Fig.~\ref{fig:s2}a. 

As an addition to our previous optical characterization setup, we also include a resonant frequency tracking setup to excite the modulator at its time-varying resonant frequency. Driving the device while being impedance matched at sufficiently high RF power levels leads to heating of the wafer, causing the resonant frequency to red shift. For sufficiently high $Q$, this can cause the modulation efficiency to drop significantly. To overcome this issue, we use a resonant frequency tracking setup, where the drive frequency for the modulator $f_r(t)$ and the modulation frequency of the AOM $f_r(t) + 4~\text{Hz}$ are adjusted simultaneously as a function of time $t$ to keep a constant beat tone at 4~Hz. We observe settling to some sort of steady-state resonant frequency after tens of seconds (thermal equilibrium), where the variation in $f_r(t)$ with respect to its mean is less than 1 in 10,000 (see Supplementary section 5 for more details). We perform characterization only after achieving this approximate steady-state in the resonant frequency.

We perform optical measurements at four different RF power levels, and extract the pixel-wise as well as rms $S'_{yz}$ for each measurement. For each measurement, 320 frames are captured with a frame rate of 32~Hz, the exposure time for each frame is 280~$\mu s$, with 16 bit precision for the pixels. The volume rms $S'_{yz}$ is calculated by using the capture without and with a polarizer placed after the modulator (P2 in Fig.~\ref{fig:s2}a). The peak-to-peak variation in the beat tone at 4~Hz when P2 is present (computed with a fast Fourier transform), and the average intensity value with and without P2 are used per pixel to calculate $S'_{yz}$. For the capture where P2 is present, the intensity as a function of time for each pixel is as shown in equation~\eqref{Eq.R3}. For the capture without P2 present, the intensity captured by each pixel is $I_0/2$. In calculating $S'_{yz}$ we use equation~\eqref{Eq.R2} and equation~\eqref{Eq.R3}, and make the following assumptions: for small $\phi_D$, $J_0(\phi_D) \approx 1$ and $J_1(\phi_D) \approx \frac{\phi_D}{2}$. We also use the fact that the depth of modulation for the laser beam is 48\% before being incident on the LN modulator. Volume rms $S'_{yz}$ is computed using these pixel-wise values.

For 160~mW of RF excitation at resonance for a 1~cm diameter region, the rms $S'_{yz}$ is $9.3 \times 10^{-5}$. Using equation~\eqref{Eq.R2}, the required rms $S'_{yz}$ to make $\phi_D = 1.2$ is $2.5 \times 10^{-4}$. The required RF power to reach 100\% intensity modulation over a $1~\text{cm}^2$ area is calculated as $0.16~\text{W} \times \Big(\frac{2.5 \times 10^{-4}}{9.3 \times 10^{-5}}\Big)^2 \times \frac{1~\text{cm}^2}{\pi \times (0.5~\text{cm})^2} = 1.5~\text{W}$. We use rms $S'_{yz}$ to report the modulation efficiency, similar to how voltage rms (spatial) amplitude is used in reference to electrical power. Compared to our previous demonstration relying on Y-cut LN with a resonant frequency of 3.7~MHz, the modulation efficiency has improved by a factor of $\big(7.4~\text{W}\text{ cm}^{-2} / 1.5~\text{W}\text{ cm}^{-2}\big) \times \big(7~\text{MHz}/3.7~\text{MHz}\big)^2 = 17.7$. $P_{RF}$ is proportional to $f_r^2$, and this can be seen by inspecting equation~\eqref{Eq.R5}. To achieve 100\% intensity modulation, the product of $L$ and rms $S'_{yz}$ should be a constant. This can be expressed as: $\frac{v'}{2f_r} \times \sqrt{\frac{\int_V S'^2_{yz}dV}{V}} = \text{constant}$, which implies that $S'_{yz}$ is proportional to $f_r$ (since $V = \pi r^2 \frac{v'}{2f_r}$). Therefore, $\int_V S'^2_{yz}dV$ is proportional to $f_r$, and thus $P_{RF}$ is proportional to $f_r^2$ (assuming 100\% intensity modulation). 

%The reason behind the modulation efficiency being related to the squared of the resonant frequency is due to needing higher strain for higher resonant frequencies (product of rms $S'_{yz}$ and $L$ should be constant), as well as its appearance in the definition of quality factor (see Supplementary section 4). 

We estimate the rms $S'_{yz}$ in the electrode region as a function of input RF power using equation~\eqref{Eq.R5}. The analytic and experimentally extracted rms $S'_{yz}$ are plotted in Fig.~\ref{fig:s2}h as Theory and Experiment, respectively. 

%However, we need to consider the impact of electrode resistance in order to fairly compare our analytical and experimental results. We see a factor of 2 discrepancy between the theory and experiments for $S'_{yz}$, which translates to a factor of 4 discrepancy in RF power. We attribute this discrepancy to the RF power coupled to the ITO electrode resistance. The effective resistance of the modulator at resonance was $11.5~\Omega$, which is the sum of the mechanical resistance of the modulator and the electrode resistance. Sheet resistance of $22~\Omega/\text{sq}$ for ITO is comparable to the total resistance of $11.5~\Omega$. Assuming only 25\% of the RF power is coupled into the acoustics, the estimated rms $S'_{yz}$ is plotted as Theory (2) in Fig.~\ref{fig:s2}h, showing a good match with the experimental results.

%These results highlight a crucial design parameter for the modulator - the electrode thickness. Choosing thin electrodes results in a large electrode resistance, whereas thick electrodes contribute to acoustic attenuation. Ideally, a cut angle for the wafer should be chosen to make the mechanical resistance and optical coupling sufficiently high. This corresponds to choosing a cut angle between $0^\circ$ and $25^\circ$ depending on the expected $Q$. We also note that the "internal efficiency" of our modulator is greater by a factor of 4 than the estimated $6.3~\text{W}/\text{cm}^2$, which implies that if the electrode resistance was negligible, the device would only require $1.6~\text{W}/\text{cm}^2$ for 100\% modulation at 7~MHz. 

We also measure the insertion loss for the modulator at 532~nm wavelength by measuring the attenuation due to having the modulator in the path of the laser beam. Using a fast photodetector, we measure the optical insertion loss to be 1.1~dB.

%%For beta = 5 degrees, p'_14 = -0.0423, p'_24 = 0.0876
%% n'_y = 2.2298

\section{Time-of-Flight Demonstration}
We demonstrate one application of our LN modulator: phase-shift based ToF imaging~\cite{CMOS_optical_ToF,pmd_ToF}. In this imaging modality, intensity modulated light is used to illuminate targets. The phase of the intensity modulated light is shifted due to the propagation of the light from the source, to the target, and back to the receiver (after reflecting from the target). The phase shift of the intensity modulation is related to the distance $d$ of a target in the scene through equation~\eqref{Eq.R6}, where $\Phi$ is the phase shift, and $c$ the speed of light in air. We use our LN modulator to down-convert the megahertz level intensity modulation $f_r$ into a hertz level beat tone that can be detected with a standard image sensor.

\begin{gather}
d = \frac{\Phi c}{4 \pi f_r} \label{Eq.R6}
\end{gather}

The modulator is excited with 160~mW of RF power with frequency $f_r(t)$ via the resonant frequency tracking setup, and a 520~nm wavelength laser diode with output power 80~mW is intensity modulated at $f_r(t) + 4$~Hz by modulating the injection current to the laser diode. We measure a depth of modulation of 72\% for the intensity modulated laser beam using a fast photodetector. The divergence of the laser beam is shaped by passing through a lens, and this beam is used to illuminate targets placed 1-2~m away from the laser. A standard CMOS camera offering 4~megapixel resolution is placed next to the laser. A camera lens with diameter 4.5~cm is attached to the camera (to resolve the targets). A polarizer - modulator - polarizer optical chain is placed in front of the camera. A pupil is placed farther from the camera, between polarizer P2 and modulator W as illustrated in Fig.~\ref{fig:s3}a so that the reflected laser light from the targets only passes through the active region of the modulator. 

%DoM for 520 nm laser estimated using 12_16_2021_ToF_Trials. Compare peak-to-peak variation to DC value. 

%A pupil is placed between polarizer (P2) and modulator (W) in Fig.~\ref{fig:s3}a so that the reflected laser light from the targets only passes through the active region of the modulator where strain is present. 

We use white wooden targets that reflect diffusely to perform a ToF imaging experiment, and place them more than a meter away from the laser and the camera (as shown in Fig.~\ref{fig:s3}). Black cardboard is placed underneath and behind the targets to limit multi-path interference (multiple light paths incident on the targets by following different paths) and to remove background light, respectively.

For the ToF measurement, 960 frames are captured with a frame rate of 16~Hz, the exposure time for each frame is 62~ms, camera gain of 10~dB is used, with 16 bit precision for the pixels. Fig.~\ref{fig:s3}c shows the brightness of the targets detected by the camera per pixel, and Fig.~\ref{fig:s3}d shows the reconstructed color coded depth per pixel (by mapping the phase at the beat tone of 4~Hz to distance using equation~\eqref{Eq.R6} after performing a fast Fourier transform across time for each pixel). Fig.~\ref{fig:s3}f shows the depth estimate distribution for the different targets. The ranging accuracy per pixel is approximately 27~cm based on the standard deviation of distance per pixel for the same target. The ranging accuracy is limited by the illumination power, RF power driving the device, the aperture of the modulator, integration time, and the resonant frequency. 

We have demonstrated ToF capability for our system on diffusely reflecting targets. Moving to higher modulation frequencies (by using thinner wafers) and improving the $Q$ through carefully choosing $\beta$ and the electrode thickness would improve the ToF performance.

\section{Conclusion}
In summary, we demonstrated Y-Z cut lithium niobate longitudinal piezoelectric resonant photoelastic modulators. These modulators consist of a simple design and allow intensity modulation of free-space beams at megahertz frequencies with record-high efficiency. The modulator described in this work allows simultaneous high modulation frequency and efficiency, and can find use in applications requiring free-space beams to be intensity modulated with low-power at megahertz frequencies. As a potential use case, we have demonstrated 4 megapixel ToF imaging relying on a standard image sensor. This work opens up the path for reaching higher frequencies (20~MHz and beyond by using thinner wafers) and with even higher modulation efficiencies. This advance would enable low-power, high performance, and high spatial resolution (exceeding hundred megapixels) ToF imaging through integrating this modulator with a standard CMOS image sensor.

\section*{Acknowledgements}
The authors thank Christopher J. Sarabalis, Felix M. Mayor, Wentao Jiang, and Prof. David A.B. Miller for useful discussions. This work was funded in part by Stanford SystemX Alliance, Office of Naval Research, and NSF ECCS-1808100.

\section*{Disclosures}
O.A., A.H.S.-N., and A.A. are inventors of US patent application 16/971,127.

\section*{Data availability}
Data underlying the results presented in this paper are not publicly available at this time but may be obtained from the corresponding author upon reasonable request.

\bibliographystyle{unsrt}
\bibliography{references}

\onecolumn
\newpage

%\centering

\begin{center}
  \section*{\textbf{\fontsize{16}{19.2}\selectfont Supplementary Material}}
\end{center} 

\bigskip \bigskip 
%\title{\textbf{\LARGE{Supplementary Information}}}
%\maketitle

\renewcommand\thefigure{S\arabic{figure}}
\setcounter{figure}{0} 

\setcounter{section}{0}

%\tableofcontents

\section{Complete Strain Profile in the Wafer}
For completeness, the other simulated strain profiles in the LN Y85 wafer when excited with 2Vpp at 6.926~MHz are shown in Fig.~\ref{fig:sup1}. For this COMSOL simulation, the quality factor is $Q = 6 \times 10^3$, the radius of the top and bottom surface electrodes is 6.35~mm (with active area radius of $r = 5~\text{mm}$), and the wafer thickness is $ L = 255 ~\mu \text{m}$. Notice that the dominant strain amplitude ($S'_{yz}$) is more than 20 times larger than any of the other strain components due to large piezoelectric coupling via the rotated piezoelectric stress tensor $e'_{34}$ component. 

%When coupled with the fact that the $S'_{yz}$ strain couples strongly to optics via the photoelastic tensor for Y-Z cut LN, this justifies the use of only this strain component for calculating the RF to optical power efficiency. 

\section{Dielectric, Elastic, and Piezoelectric Properties of Lithium Niobate}
The modulator is made of lithium niobate, which requires us to use the relevant properties of this material to determine the modulator characteristics. The material properties are fully described in the linear regime by the relevant tensors. The dielectric, elastic, and piezoelectric tensors for lithium niobate (belonging to trigonal crystal system with point group 3m) are shown below~\cite{auld_vol1}, where $\mathbf{\epsilon}$ is the dielectric tensor, $\mathbf{S}$ the compliance tensor, $\mathbf{C}$ the stiffness tensor, $\mathbf{d}$ the piezoelectric strain tensor, $\mathbf{e}$ the piezoelectric stress tensor:

\begin{gather}
\mathbf{\epsilon} = 
\begin{pmatrix}
\epsilon_{11} & 0 & 0 \\
0 & \epsilon_{11} & 0 \\
0 & 0 & \epsilon_{33} \label{Eq.1} \tag{S1}
\end{pmatrix} 
\end{gather}

\begin{gather}
\mathbf{S} = 
\begin{pmatrix}
s_{11} & s_{12} & s_{13} & s_{14} & 0 & 0  \\
s_{12} & s_{11} & s_{13} & -s_{14} & 0 & 0 \\
s_{13} & s_{13} & s_{33} & 0 & 0 & 0 \\
s_{14} & -s_{14} & 0 & s_{44} & 0 & 0 \\
0 & 0 & 0 & 0 & s_{44} & 2s_{14} \\
0 & 0 & 0 & 0 & 2s_{14} & 2(s_{11} - s_{12}) \label{Eq.2} \tag{S2}
\end{pmatrix} 
\end{gather}

\begin{gather}
\mathbf{C} = 
\begin{pmatrix}
c_{11} & c_{12} & c_{13} & c_{14} & 0 & 0  \\
c_{12} & c_{11} & c_{13} & -c_{14} & 0 & 0 \\
c_{13} & c_{13} & c_{33} & 0 & 0 & 0 \\
c_{14} & -c_{14} & 0 & c_{44} & 0 & 0 \\
0 & 0 & 0 & 0 & c_{44} & c_{14} \\
0 & 0 & 0 & 0 & c_{14} & 0.5(c_{11} - c_{12}) \label{Eq.3} \tag{S3}
\end{pmatrix} 
\end{gather}

\begin{gather}
\mathbf{d} = 
\begin{pmatrix}
0 & 0 & 0 & 0 & d_{15} & -2d_{22}  \\
-d_{22} & d_{22} & 0 & d_{15} & 0 & 0 \\
d_{31} & d_{31} & d_{33} & 0 & 0 & 0 \label{Eq.4} \tag{S4}
\end{pmatrix} 
\end{gather}

\begin{gather}
\mathbf{e} = 
\begin{pmatrix}
0 & 0 & 0 & 0 & e_{15} & -e_{22}  \\
-e_{22} & e_{22} & 0 & e_{15} & 0 & 0 \\
e_{31} & e_{31} & e_{33} & 0 & 0 & 0 \label{Eq.5} \tag{S5}
\end{pmatrix} 
\end{gather}

To work in the rotated coordinate frame (x',y',z') to simplify the analysis, the Bond transformation matrices are applied to the crystal coordinate frame (x,y,z). The Bond transformation matrices and the operations needed to transform from (x,y,z) to the rotated coordinate frame (x',y',z') are given below, expressed in primed notation:

\begin{figure*}[t!]
\centering
\includegraphics[width=1\textwidth]{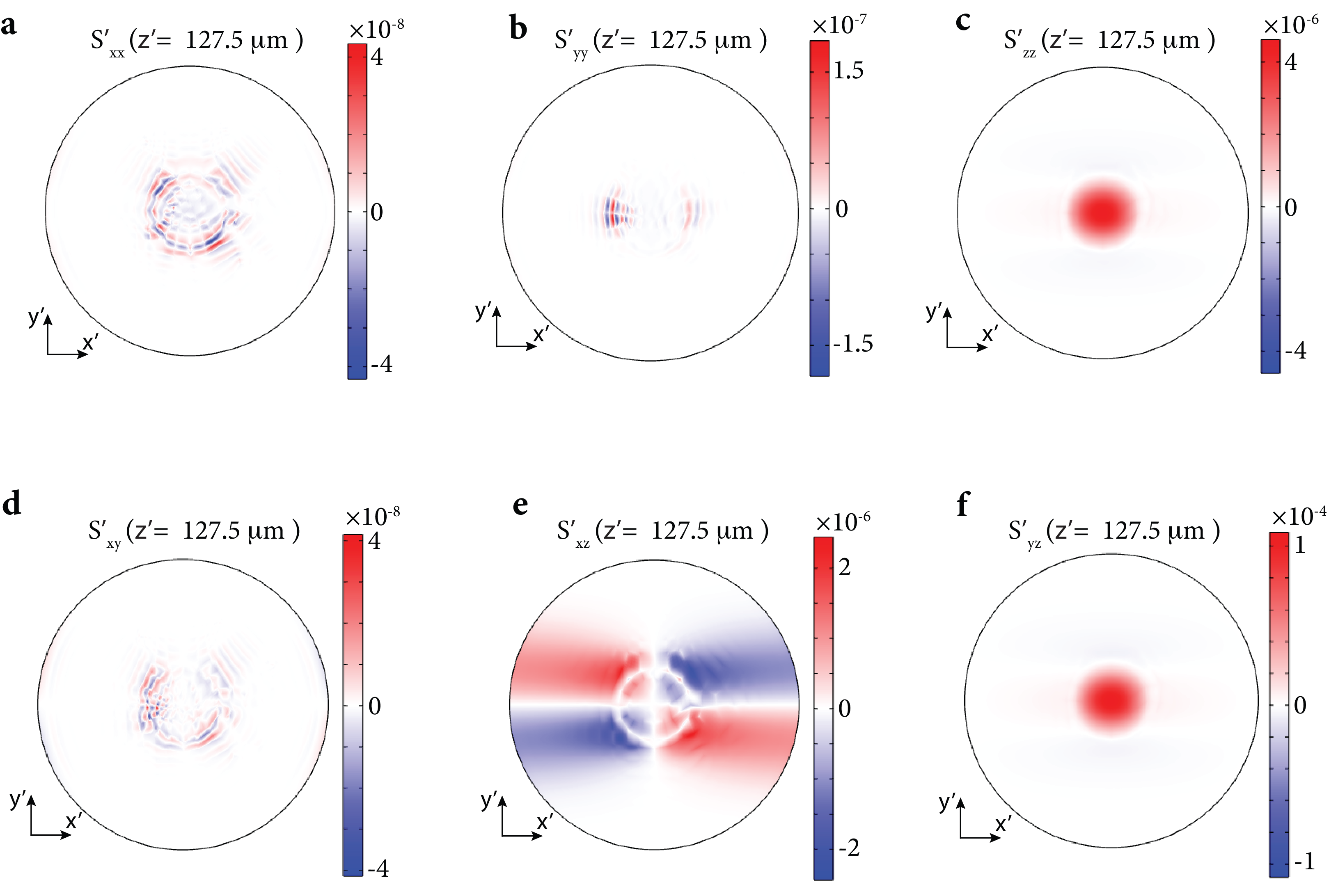}
\caption{Strain amplitude profiles in the wafer. (a) $S'_{xx}$ strain amplitude profile in the rotated coordinate frame when the wafer is excited at 6.926~MHz with 2Vpp applied to the surface electrodes is shown for the center of the wafer. (b) $S'_{yy}$ strain amplitude profile in the rotated coordinate frame when the wafer is excited at 6.926~MHz with 2Vpp applied to the surface electrodes is shown for the center of the wafer. (c) $S'_{zz}$ strain amplitude profile in the rotated coordinate frame when the wafer is excited at 6.926~MHz with 2Vpp applied to the surface electrodes is shown for the center of the wafer. (d) $S'_{xy}$ strain amplitude profile in the rotated coordinate frame when the wafer is excited at 6.926~MHz with 2Vpp applied to the surface electrodes is shown for the center of the wafer. (e) $S'_{xz}$ strain amplitude profile in the rotated coordinate frame when the wafer is excited at 6.926~MHz with 2Vpp applied to the surface electrodes is shown for the center of the wafer. (f) The dominant $S'_{yz}$ strain amplitude profile in the rotated coordinate frame when the wafer is excited at 6.926~MHz with 2Vpp applied to the surface electrodes is shown for the center of the wafer.}
\label{fig:sup1}
\end{figure*}

\begin{gather}
\mathbf{R} = 
\begin{pmatrix}
1 & 0 & 0 \\
0 & \text{cos}\beta & \text{sin}\beta \\
0 & -\text{sin}\beta & \text{cos}\beta \label{Eq.6} \tag{S6}
\end{pmatrix} 
\end{gather}

\begin{gather}
\mathbf{M} = 
\begin{pmatrix}
1 & 0 & 0 & 0 & 0 & 0  \\
0 & \text{cos}^2 \beta & \text{sin}^2 \beta & 2 \text{cos} \beta \text{sin} \beta & 0 & 0 \\
0 & \text{sin}^2 \beta & \text{cos}^2 \beta & -2 \text{cos} \beta \text{sin} \beta & 0 & 0 \\
0 & -\text{cos} \beta \text{sin} \beta & \text{cos} \beta \text{sin} \beta & 1 & 0 & 0 \\
0 & 0 & 0 & 0 & \text{cos} \beta & -\text{sin} \beta \\ 
0 & 0 & 0 & 0 & \text{sin} \beta & \text{cos} \beta \label{Eq.7} \tag{S7}
\end{pmatrix} 
\end{gather}

\begin{gather}
\mathbf{N} = 
\begin{pmatrix}
1 & 0 & 0 & 0 & 0 & 0  \\
0 & \text{cos}^2 \beta & \text{sin}^2 \beta & \text{cos} \beta \text{sin} \beta & 0 & 0 \\
0 & \text{sin}^2 \beta & \text{cos}^2 \beta & -\text{cos} \beta \text{sin} \beta & 0 & 0 \\
0 & -2\text{cos} \beta \text{sin} \beta & 2\text{cos} \beta \text{sin} \beta & 1 & 0 & 0 \\
0 & 0 & 0 & 0 & \text{cos} \beta & -\text{sin} \beta \\ 
0 & 0 & 0 & 0 & \text{sin} \beta & \text{cos} \beta \label{Eq.8} \tag{S8}
\end{pmatrix} 
\end{gather}

\begin{gather}
\mathbf{\epsilon'} = \mathbf{R}\mathbf{\epsilon}\mathbf{R}^T \label{Eq.9} \tag{S9}
\end{gather}

\begin{gather}
\mathbf{S'} = \mathbf{N}\mathbf{S}\mathbf{N}^T \label{Eq.10} \tag{S10}
\end{gather}

\begin{gather}
\mathbf{C'} = \mathbf{M}\mathbf{C}\mathbf{M}^T \label{Eq.11} \tag{S11}
\end{gather}

\begin{gather}
\mathbf{d'} = \mathbf{R}\mathbf{d}\mathbf{N}^T \label{Eq.12} \tag{S12}
\end{gather}

\begin{gather}
\mathbf{e'} = \mathbf{R}\mathbf{e}\mathbf{M}^T \label{Eq.13} \tag{S13}
\end{gather}

In the primed tensor notation, $w'_{ij}$ denotes the element in the i\textsuperscript{th} row and j\textsuperscript{th} column of the rotated $\mathbf{W'}$ tensor. $\mathbf{\epsilon'}$ is the rotated dielectric tensor, $\mathbf{S'}$ the rotated compliance tensor, $\mathbf{C'}$ the rotated stiffness tensor, $\mathbf{d'}$ the rotated piezoelectric strain tensor, $\mathbf{e'}$ the rotated piezoelectric stress tensor. 

\subsection{Solving the Christoffel Equation}
To determine the resonant frequency of the desired $S'_{yz}$ shear mode for a given thickness $L$ for the wafer, we need to calculate the phase velocity for this mode. We are interested in the phase velocity of the $S'_{yz}$ acoustic wave propagating along the z' direction of the wafer. The phase velocity for this mode is found (approximately) by solving the Christoffel equation shown in equation \eqref{Eq.14}. In this equation, $\rho$ is the density of the material, $v_p$ the phase velocity of the corresponding mode, $\mathbf{v}$ the polarization vector for the acoustic field, and $\mathbf{\Gamma}$ the Christoffel matrix. The elements of the Christoffel matrix are given in equation \eqref{Eq.15}, where $c_{ij}$ are the stiffness coefficients of lithium niobate ($c_{66} = \frac{1}{2}(c_{11} - c_{12})$). This gives three different acoustic fields that propagate along z'; we choose the one with the largest transverse component, with phase velocity $v'$. The phase velocity of this field is related to the resonant frequency of the modulator (plane wave assumption) by $f_r = \frac{v'}{2L}$.

\begin{gather}
\frac{\mathbf{\Gamma}}{\rho}\mathbf{v} = v_p^2\mathbf{v} \label{Eq.14} \tag{S14}
\end{gather}

\begin{gather}
\mathbf{\Gamma} = 
\begin{pmatrix}
c_{66}\text{sin}^2\beta + c_{44}\text{cos}^2\beta + 2c_{14}\text{cos}\beta \text{sin} \beta & 0 & 0 \\
0 & c_{11}\text{sin}^2\beta + c_{44}\text{cos}^2\beta - 2c_{14}\text{cos}\beta \text{sin} \beta & (c_{13} + c_{44})\text{cos}\beta \text{sin} \beta - c_{14}\text{sin}^2\beta \\
0 & (c_{13} + c_{44})\text{cos}\beta \text{sin} \beta - c_{14}\text{sin}^2\beta  & c_{44}\text{sin}^2\beta + c_{33}\text{cos}^2\beta \label{Eq.15} \tag{S15}
\end{pmatrix} 
\end{gather}

\section{BVD Model of the Modulator}
\begin{figure*}[t!]
\centering
\includegraphics[width=0.75\textwidth]{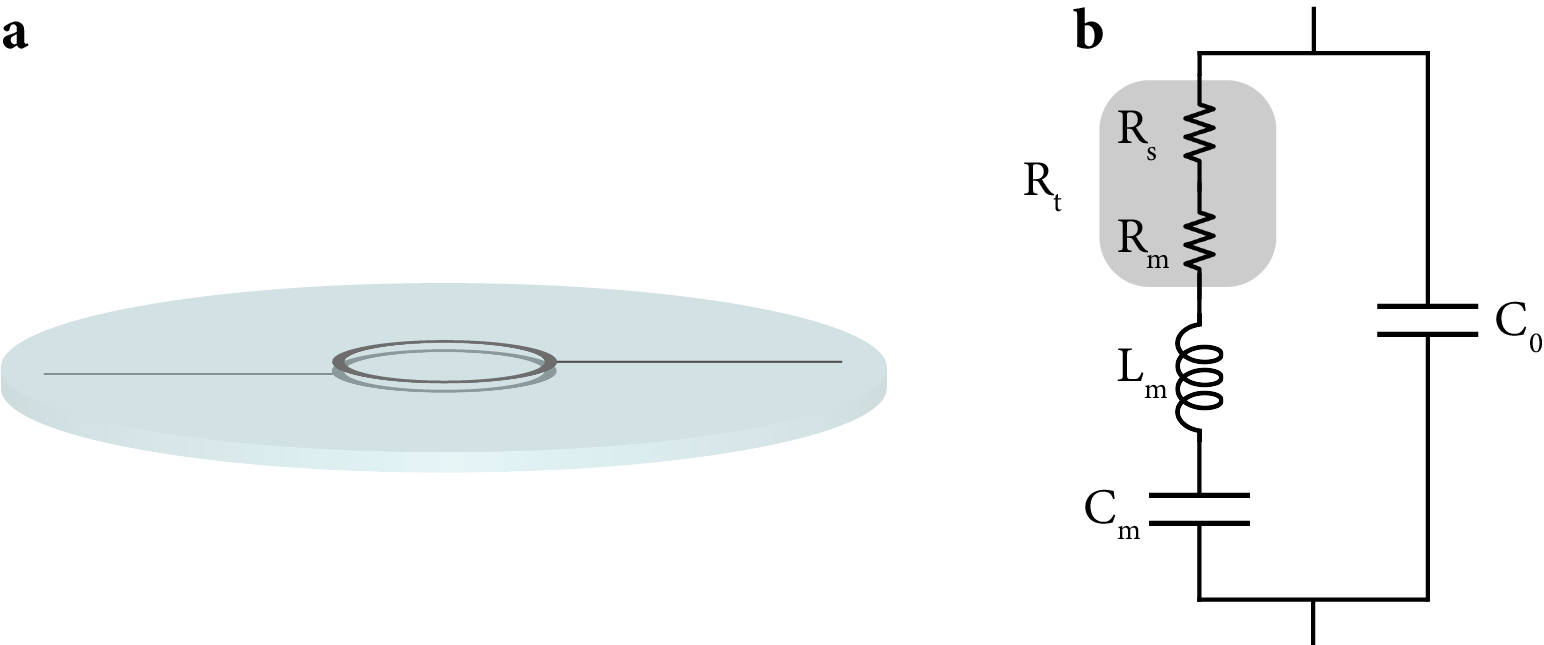}
\caption{BVD equivalent circuit of the piezoelectric resonator. (a) Lithium niobate wafer coated on top and bottom surfaces with electrodes. (b) BVD equivalent circuit for the $S'_{yz}$ shear mode of the wafer shown in (a).}
\label{fig:sup2}
\end{figure*}

Piezoelectric modulators can be modeled using the Butterworth Van-Dyke (BVD) model. One mode of the modulator can be modeled in the electrical domain using a single branch BVD model; a mechanical RLC branch in parallel with an electrical capacitance. The RLC branch models the linear piezoelectricity, and fully defines the electromechanical behavior. The equivalent BVD model for the modulator is shown in Fig.~\ref{fig:sup2}. 

In this section, we will derive the relevant parameters with $L_m$, $C_m$, and $R_{t}$ ($R_t = R_m + R_s$, where $R_m$ is the mechanical resistance and $R_s$ the lumped electrode resistance) constituting the mechanical branch and $C_0$ constituting the parallel capacitance. This analysis assumes that $S'_{yz}$ is the only strain profile present in the wafer. We make this assumption to simplify the analysis and expressions and because $S'_{yz}$ is the dominant strain profile in the wafer (as shown in Fig.~\ref{fig:sup1}). Our starting point is the linear piezoelectric constitutive equations, expressed below:

\begin{gather}
S'_{yz} = d'_{34}E'_z + s'_{44}T'_{yz} \nonumber \\
T'_{yz} = -e'_{34}E'_z + c'_{44}S'_{yz} \label{Eq.16} \tag{S16}
\end{gather}

For the equations above, $T'_{yz}$ is the shear stress amplitude in the rotated coordinate frame, and $E'_z$ the electric field amplitude due to the RF voltage $V$ on the surface electrodes of the wafer in the rotated coordinate frame. Substituting the second equation into the first, we get:

\begin{gather}
S'_{yz} = d'_{34}E'_z + s'_{44}(-e'_{34}E'_z + c'_{44}S'_{yz}) \label{Eq.17} \tag{S17}
\end{gather}

Using $E'_z\text{cos}(2 \pi f_r t) = \frac{V}{L} = \frac{2Vf_r}{v'}$, the above expression can be expressed as:

\begin{gather}
S'_{yz}\text{cos}(2 \pi f_r t) = \frac{2Vf_r(d'_{34} - s'_{44}e'_{34})}{v'(1 - c'_{44}s'_{44})} \label{Eq.18} \tag{S18}
\end{gather}

For the expression above, $V = V_0 \text{cos}(2 \pi f_r t)$, where $V_0$ is the amplitude and $t$ the time. We now relate $S'_{yz}$ to the electric displacement through  $D = e'_{34}S'_{yz}\text{cos}(2 \pi f_r t)$. This only includes the piezoelectric contribution (does not include electrical capacitance contribution), since we are interested in finding $C_m$:

\begin{gather}
I = \int_{S} \frac{d}{dt} D dS = C_m\frac{dV}{dt} =  \pi r^2 \frac{e'_{34}(d'_{34} - s'_{44}e'_{34}) 4 \pi V f_r^2}{v'(1 - c'_{44}s'_{44})} \label{Eq.19} \tag{S19}
\end{gather}

For the expression above, $I$ is the displacement current and the integral is carried out over the surface electrodes (with radius $r$). We can now determine all the components in the BVD model as shown below:

\begin{gather}
C_m = \pi r^2 \frac{2f_re'_{34}(d'_{34} - s'_{44}e'_{34})}{v'(1 - c'_{44}s'_{44})} \approx \pi r^2 \frac{2f_re'^2_{34}}{c'_{44}v'} \label{Eq.20} \tag{S20}
\end{gather}

\begin{gather}
L_m = \frac{1}{(2 \pi f_r)^2C_m} = \frac{(1 - c'_{44}s'_{44})v'}{8f_r^3 \pi^3 r^2 (d'_{34} - s'_{44}e'_{34})e'_{34}} \approx \frac{c'_{44}v'}{8f_r^3 \pi^3 r^2 e'^2_{34}} \label{Eq.21} \tag{S21}
\end{gather}

\begin{gather}
R_t = \frac{2 \pi f_r L_m}{Q} = \frac{(1 - c'_{44}s'_{44})v'}{4f_r^2 \pi^2 r^2 (d'_{34} - s'_{44}e'_{34})e'_{34}Q} \approx \frac{c'_{44}v'}{4f_r^2 \pi^2 r^2 e'^2_{34}Q} \label{Eq.22} \tag{S22}
\end{gather}

\begin{gather}
C_0 = \frac{2f_r\epsilon'_{33}\pi r^2}{v'} \label{Eq.23} \tag{S23}
\end{gather}

This model proves to be accurate if the $S'_{yz}$ is significantly larger than the other strain components. This assumption is true for $\beta$ close to 0 and $\frac{\pi}{2}$ radians, but due to the longitudinal $S'_{zz}$ component becoming comparable to the $S'_{yz}$ for $\beta$ around $\frac{\pi}{4}$, we see a discrepancy in the estimate and the COMSOL simulation (see Fig. 1g). 

%We also note that in practice, the extracted resistance $R_m$ is larger than the expression given in equation~\eqref{Eq.22} due to non-zero electrode resistance. The extracted resistance is the sum of $R_m$ (given in equation~\eqref{Eq.22}) and the electrode resistance. 

\section{RF Power Required for Optical Modulation}
To calculate the RF power required for a certain optical modulation level, we need to take into account RF to acoustic transduction via the piezoelectric tensor, and the acoustic to optical coupling via the photoelastic tensor. Assuming the RF source is impedance matched to $R_t$, we will calculate the required RF power to reach some root mean square (rms) $S'_{yz}$ strain level for the electrode region in the wafer. This calculation places a lower bound on the required RF power to operate the modulator. 

We will work backwards here and determine the required RF power to achieve a desired level of intensity modulation at steady-state operation. The intensity of a plane wave with initial intensity $I_0$ that has propagated through the polarizer, wafer, polarizer with perpendicular incidence is expressed as follows (see~\cite{longitudinal_piezo_photo_mod} for derivation):

\begin{gather}
I(t) = \frac{I_0}{2}\Big(\frac{1}{2} + \frac{1}{2}\big[\text{cos}(\phi_s)(J_0(\phi_D) - 2\text{sin}(\phi_s)J_{1}(\phi_D)\text{cos}(2 \pi f_r t) + \text{HOH}\big]\Big) \label{Eq.24} \tag{S24}
\end{gather}

For equation \eqref{Eq.24}, $J_0$ and $J_1$ are the zeroth and first order Bessel function of the first kind, respectively; HOH stands for the higher order harmonics. $\phi_s$ and $\phi_D$ are the static and dynamic polarization rotations, respectively. To determine $\phi_s$ and $\phi_D$, we use the rotated index ellipsoid in the wafer. The index ellipsoid in the rotated frame is expressed approximately as follows (only taking $S'_{yz}$ contribution)~\cite{longitudinal_piezo_photo_mod,crystal_optics}:

\begin{gather}
x^2\Bigg(\frac{1}{n_o^2} + 2p'_{14}\sqrt{\frac{\int_V S'^2_{yz}dV}{\pi r^2 L}}\text{cos}(2 \pi f_r t)\Bigg) + y'^2\Bigg(\frac{1}{{n'_y}^2} + 2p'_{24}\sqrt{\frac{\int_V S'^2_{yz}dV}{\pi r^2 L}}\text{cos}(2 \pi f_r t)\Bigg) \nonumber \\  + z'^2\Bigg(\frac{1}{{n'_z}^2} + 2p'_{34}\sqrt{\frac{\int_V S'^2_{yz}dV}{\pi r^2 L}}\text{cos}(2 \pi f_r t)\Bigg) + 2 y' z'\Bigg(2p'_{44}\sqrt{\frac{\int_V S'^2_{yz}dV}{\pi r^2 L}}\text{cos}(2 \pi f_r t)\Bigg) = 1  \label{Eq.25} \tag{S25}
\end{gather}

Where $\frac{1}{{n'_x}^2(t)} = \frac{1}{n_o^2} + 2p'_{14}\sqrt{\frac{\int_V S'^2_{yz}dV}{\pi r^2 L}}\text{cos}(2 \pi f_r t)$ with $n_o$ being the ordinary refractive index of lithium niobate and $\frac{1}{{n'_y}^2(t)} = \frac{1}{{n'_y}^2} + 2p'_{24}\sqrt{\frac{\int_V S'^2_{yz}dV}{\pi r^2 L}}\text{cos}(2 \pi f_r t)$. The rotated photoelastic tensor components are found as follows, where $\mathbf{P}$ is the photoelastic tensor and $\mathbf{P'}$ the rotated photoelastic tensor:

\begin{gather}
\mathbf{P} = 
\begin{pmatrix}
p_{11} & p_{12} & p_{13} & p_{14} & 0 & 0  \\
p_{12} & p_{11} & p_{13} & -p_{14} & 0 & 0 \\
p_{31} & p_{31} & p_{33} & 0 & 0 & 0 \\
p_{41} & -p_{41} & 0 & p_{44} & 0 & 0 \\
0 & 0 & 0 & 0 & p_{44} & p_{41} \\
0 & 0 & 0 & 0 & p_{14} & (p_{11} - p_{12})/2 \label{Eq.26} \tag{S26} 
\end{pmatrix} 
\end{gather}

\begin{gather}
\mathbf{P'} = \mathbf{M}\mathbf{P}\mathbf{M}^T \label{Eq.27} \tag{S27}
\end{gather}

The time-varying phase of the ordinary and extraordinary wave excited in the modulator are expressed as follows:

\begin{gather}
\phi_o(t) = \frac{2 \pi L}{\lambda}n'_x(t) \label{Eq.28} \tag{S28}
\end{gather}

\begin{gather}
\phi_e(t) = \frac{2 \pi L}{\lambda}n'_y(t) \label{Eq.29} \tag{S29}
\end{gather}

%We can express the static ($\phi_s$) and dynamic ($\phi_D$) phase difference (between the two polarization components) as: 

Using $\phi_s + \phi_D \text{cos}(2 \pi f_r t) = \phi_o(t) - \phi_e(t)$, $\phi_s$ and $\phi_D$ are found as:

%\begin{gather}
%\phi_s + \phi_D \text{cos}(2 \pi f_r t) = \phi_o(t) - \phi_e(t) \label{Eq.30} \tag{S30}
%\end{gather}

\begin{gather}
\phi_s = \frac{2 \pi L(n_o - n_y')}{\lambda} \label{Eq.30} \tag{S30}
\end{gather}

\begin{gather}
\phi_D = \frac{2 \pi L}{\lambda} \Big(n_o^3 p'_{14} - {n'_y}^3 p'_{24}\Big)\sqrt{\frac{\int_V S'^2_{yz}dV}{\pi r^2 L}} \label{Eq.31} \tag{S31}
\end{gather}

We will now relate the strain level amplitude $S'_{yz}$ in the wafer to the required RF power to sustain a certain optical intensity modulation level. The total stored energy in the wafer is the sum of the acoustic and electrical energy. However, we note that the acoustic energy is orders of magnitude larger than the electrical energy stored via the electrical capacitance. Therefore, we will only consider the total acoustic energy stored. The total acoustic energy stored $E_{acoustic}$ (taking only $S'_{yz}$ into consideration) is given as:

\begin{gather}
E_{acoustic} = \int_{V} \frac{1}{2}\rho v^2 dV \label{Eq.32} \tag{S32}
\end{gather}

For the equation above, $v$ denotes the amplitude of the particle velocity, and the integral is carrier out over the volume corresponding to the electrode region (with volume $\pi r^2 L)$. Since acoustic to optical coupling is governed by strain, we would like to express the total stored energy in terms of $S'_{yz}$. This is related to $S'_{yz}$ via $T'_{yz} = 2c'_{44}S'_{yz}$ and $T'_{yz} = \rho v' v$. The total stored energy can now be expressed as:

\begin{gather}
E_{acoustic} = \int_{V} \frac{2c'^2_{44}S'^2_{yz}}{\rho v'^2} dV = \int_{V} 2c'_{44}S'^2_{yz} dV \label{Eq.33} \tag{S33}
\end{gather} 

Using the definition of the quality factor $Q$ ($Q = 2\pi f_r\frac{E_{acoustic}}{P_{RF}}$), the RF power required to operate the device at the desired rms strain $S'_{yz}$ level is found to be:

\begin{gather} 
P_{RF} = \frac{4 \pi f_r c'_{44} \int_V S'^2_{yz}dV}{Q} \label{Eq.34} \tag{S34}
\end{gather}

\begin{figure*}[t!]
\centering
\includegraphics[width=1\textwidth]{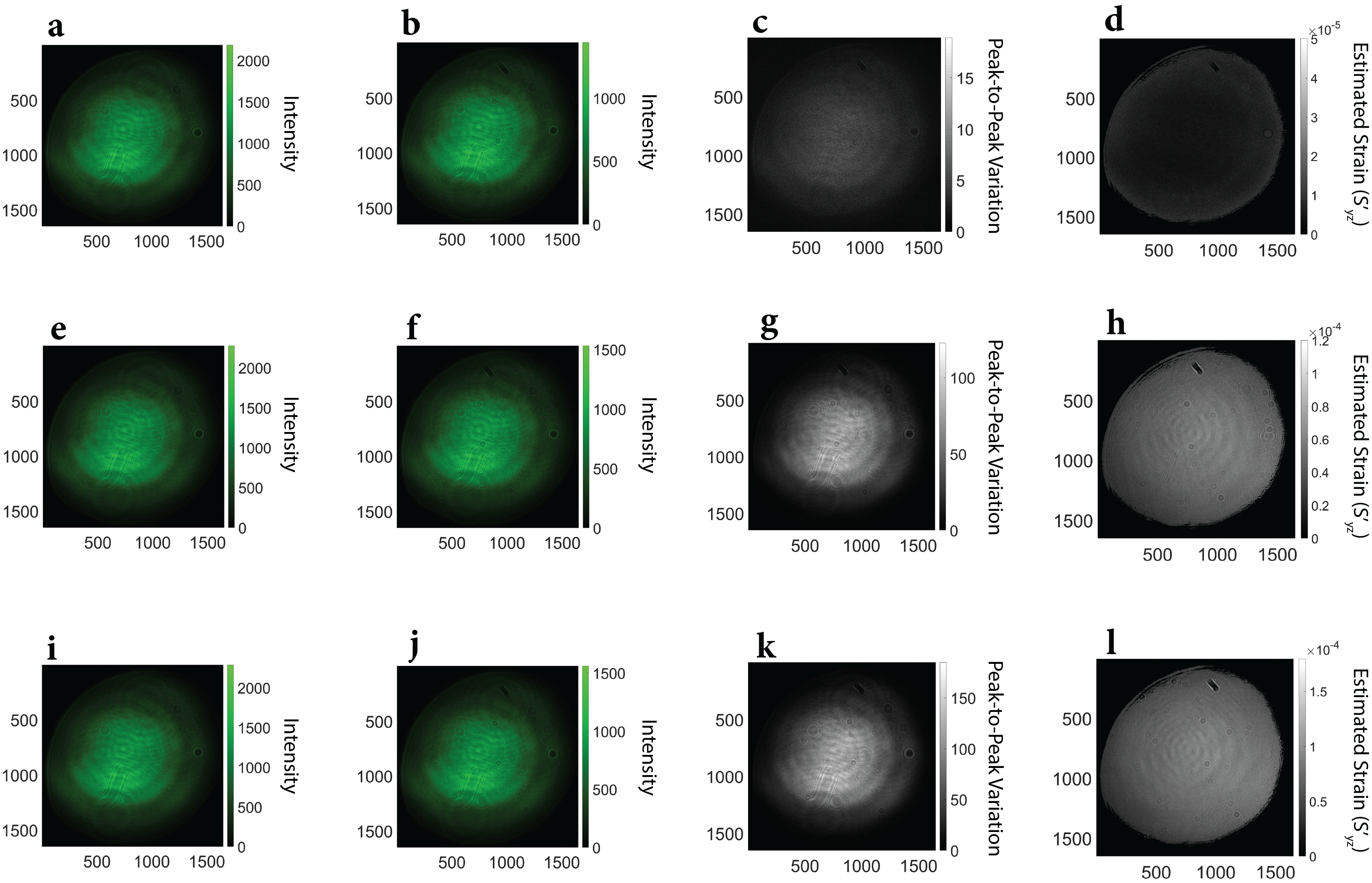}
\caption{Optical characterization for varying levels of RF excitation power. (a) Time-averaged intensity profile of the laser beam detected by the camera per pixel is shown when 0.63~mW of RF power at $f_r(t)$ is applied to the modulator and the second polarizer P2 is removed. (b) Time-averaged intensity profile of the laser beam detected by the camera per pixel is shown when 0.63~mW of RF power at $f_r(t)$ is applied to the modulator and the second polarizer P2 is present. (c) Peak-to-peak variation of the beat tone at 4~Hz of the laser beam detected by the camera per pixel is shown when 0.63~mW of RF power at $f_r(t)$ is applied to the modulator and the second polarizer P2 is present. (d) Estimated strain amplitude per pixel is shown when 0.63~mW of RF power at $f_r(t)$ is applied to the modulator and the second polarizer P2 is present. (e) Time-averaged intensity profile of the laser beam detected by the camera per pixel is shown when 40~mW of RF power at $f_r(t)$ is applied to the modulator and the second polarizer P2 is removed. (f) Time-averaged intensity profile of the laser beam detected by the camera per pixel is shown when 40~mW of RF power at $f_r(t)$ is applied to the modulator and the second polarizer P2 is present. (g) Peak-to-peak variation of the beat tone at 4~Hz of the laser beam detected by the camera per pixel is shown when 40~mW of RF power at $f_r(t)$ is applied to the modulator and the second polarizer P2 is present. (h) Estimated strain amplitude per pixel is shown when 40~mW of RF power at $f_r(t)$ is applied to the modulator and the second polarizer P2 is present. (i) Time-averaged intensity profile of the laser beam detected by the camera per pixel is shown when 90~mW of RF power at $f_r(t)$ is applied to the modulator and the second polarizer P2 is removed. (j) Time-averaged intensity profile of the laser beam detected by the camera per pixel is shown when 90~mW of RF power at $f_r(t)$ is applied to the modulator and the second polarizer P2 is present. (k) Peak-to-peak variation of the beat tone at 4~Hz of the laser beam detected by the camera per pixel is shown when 90~mW of RF power at $f_r(t)$ is applied to the modulator and the second polarizer P2 is present. (l) Estimated strain amplitude per pixel is shown when 90~mW of RF power at $f_r(t)$ is applied to the modulator and the second polarizer P2 is present.}
\label{fig:sup3}
\end{figure*}

\begin{figure*}[t!]
\centering
\includegraphics[width=1\textwidth]{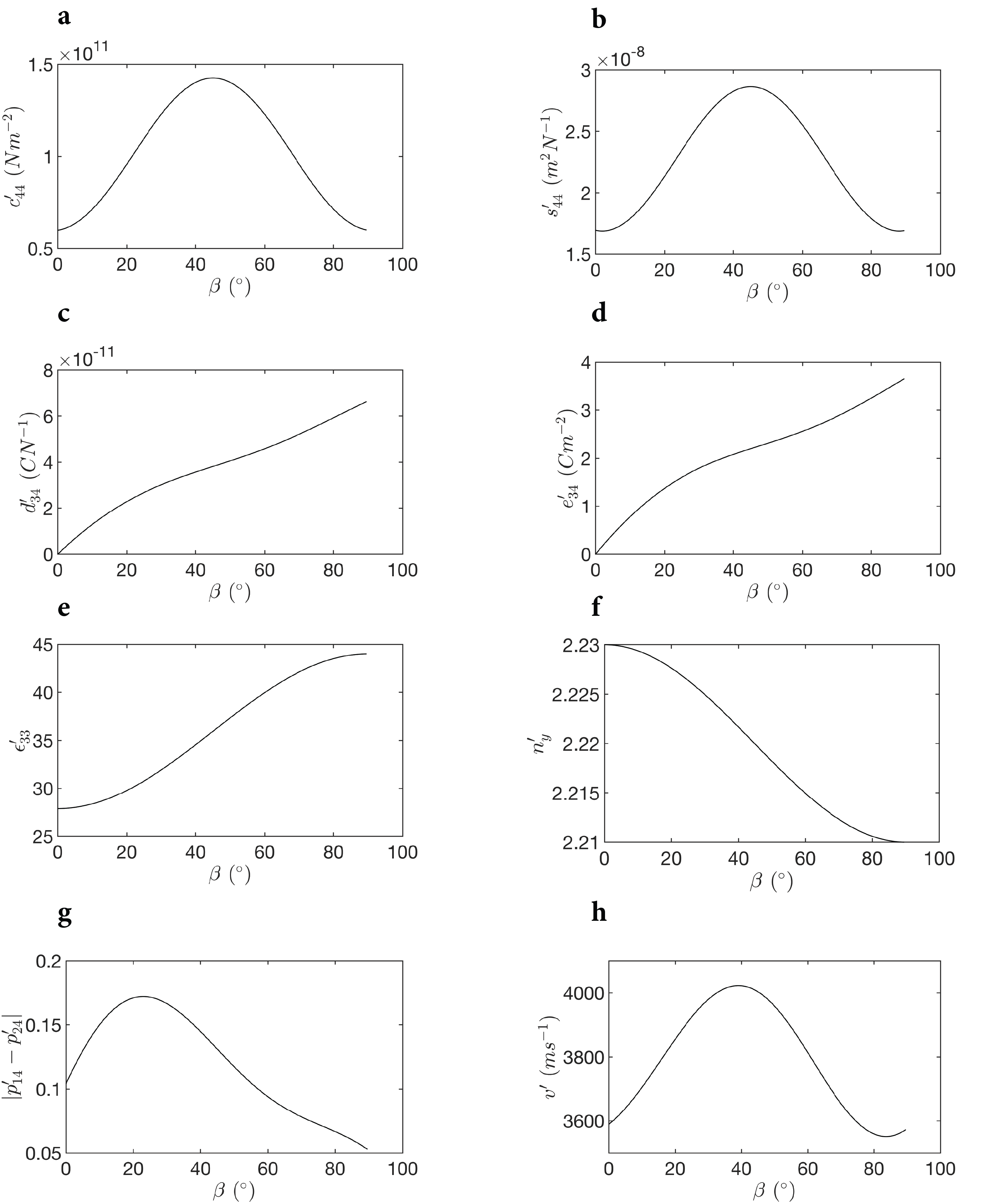}
\caption{Variation of relevant dielectric, elastic, piezoelectric, and photoelastic parameters as a function of the cut angle $\beta$. (a) Stiffness coefficient $c'_{44}$ in the primed coordinate frame as a function of $\beta$ is shown. (b) Compliance coefficient $s'_{44}$ in the primed coordinate frame as a function of $\beta$ is shown. (c) Piezoelectric strain coefficient $d'_{34}$ in the primed coordinate frame as a function of $\beta$ is shown. (d) Piezoelectric stress coefficient $e'_{34}$ in the primed coordinate frame as a function of $\beta$ is shown. (e) Relative permittivity $\epsilon'_{33}$ in the primed coordinate frame as a function of $\beta$ is shown. (f) Refractive index $n'_{y}$ in the primed coordinate frame as a function of $\beta$ is shown. (g) Absolute difference of the relevant photoelastic coefficients $|p'_{14} - p'_{24}|$ in the primed coordinate frame as a function of $\beta$ is shown. (h) Speed of propagation $v'$ for the mode with dominant shear strain $S'_{yz}$ in the primed coordinate frame as a function of $\beta$ is shown.}
\label{fig:sup4}
\end{figure*}

For completeness, optical characterization results at different RF excitation power levels are shown in Fig.~\ref{fig:sup3}. Additionally, the variation of the relevant dielectric, elastic, piezoelectric, and photoelastic properties are shown in Fig.~\ref{fig:sup4} as a function of $\beta$ using the values reported in~\cite{photoelastic_constants}. 

\section{Resonant Frequency Tracking}
The resonant frequency of the modulator is tracked using the setup shown in Fig. 2b. The two channels of the oscilloscope are terminated with 50~$\Omega$ impedances and record the RF voltage waveforms sent into and reflected from the device, respectively. These two waveforms are inputs to a MATLAB script that estimates the time-varying resonant frequency of the modulator and adjusts the drive frequency accordingly. The MATLAB script operates in two phases: 1) a crude frequency stabilization executed prior to each imaging experiment and 2) a fine frequency negative feedback control mode that operates during imaging experiments. A new crude frequency stabilization phase is executed after changes in modulator drive power, including any extended idle periods.

The pre-experiment crude frequency stabilization serves to 'burn-in' the modulator by increasing its temperature to an approximate steady state for a given RF power level. This 'burn-in' is achieved by driving the modulator for approximately 100~s with a fixed voltage amplitude and an updating frequency that crudely approximates the resonant frequency using reflected waveform magnitude measurements. The drive frequency is updated using the following algorithm, which is looped through 40 times.
\begin{enumerate}
    \item The modulator is excited with a fixed frequency $f_r$ for 2.5~s. For the initial iteration, this frequency is the resonant frequency obtained from low power (0~dBm) VNA measurements.
    \item The drive frequency is quickly swept with $f_r$ as a center frequency over a 1~kHz span with 50~Hz spacing.  A single reflected RF waveform is recorded for each frequency in the sweep.
    \item The sweep frequency with the lowest reflected RF power is identified as the best approximation of the current resonant frequency. This sweep frequency is selected as $f_r$ for the next iteration of the crude stabilization loop. 
\end{enumerate}
During the crude frequency stabilization phase, we observe that $f_r$ initially decreases with each loop iteration due to self-heating of the wafer. After tens of seconds, we note that the device reaches an approximate steady-state, with variations in $f_r$ spanning some center frequency.

After completion of the crude frequency stabilization phase, the fine frequency control phase is initiated. In fine frequency control, a negative feedback loop is used to adjust the drive frequency $f_r(t)$ to minimize $|s_{11}|$ for the modulator. Phase difference between the modulator's reflected and input RF waveforms serves as the error term in this negative feedback control scheme. Phase difference minimization is a convenient proxy for reflected power minimization given the available instrumentation and power levels. As an error term, phase enables resolution of the required sign of drive frequency updates, which cannot be deduced reliably from magnitude measurements alone. The negative feedback controller is of the proportional-integral type and implemented via software in the MATLAB script using manually tuned gain values. Proportional gain constant is 0.2~Hz/deg. Integral gain constant is 0.0625~Hz/deg. $f_r(t)$ is updated via this software controller approximately every 150~ms, a rate set by the speed of waveform capture and subsequent processing.

\begin{figure*}[t!]
\centering
\includegraphics[width=1\textwidth]{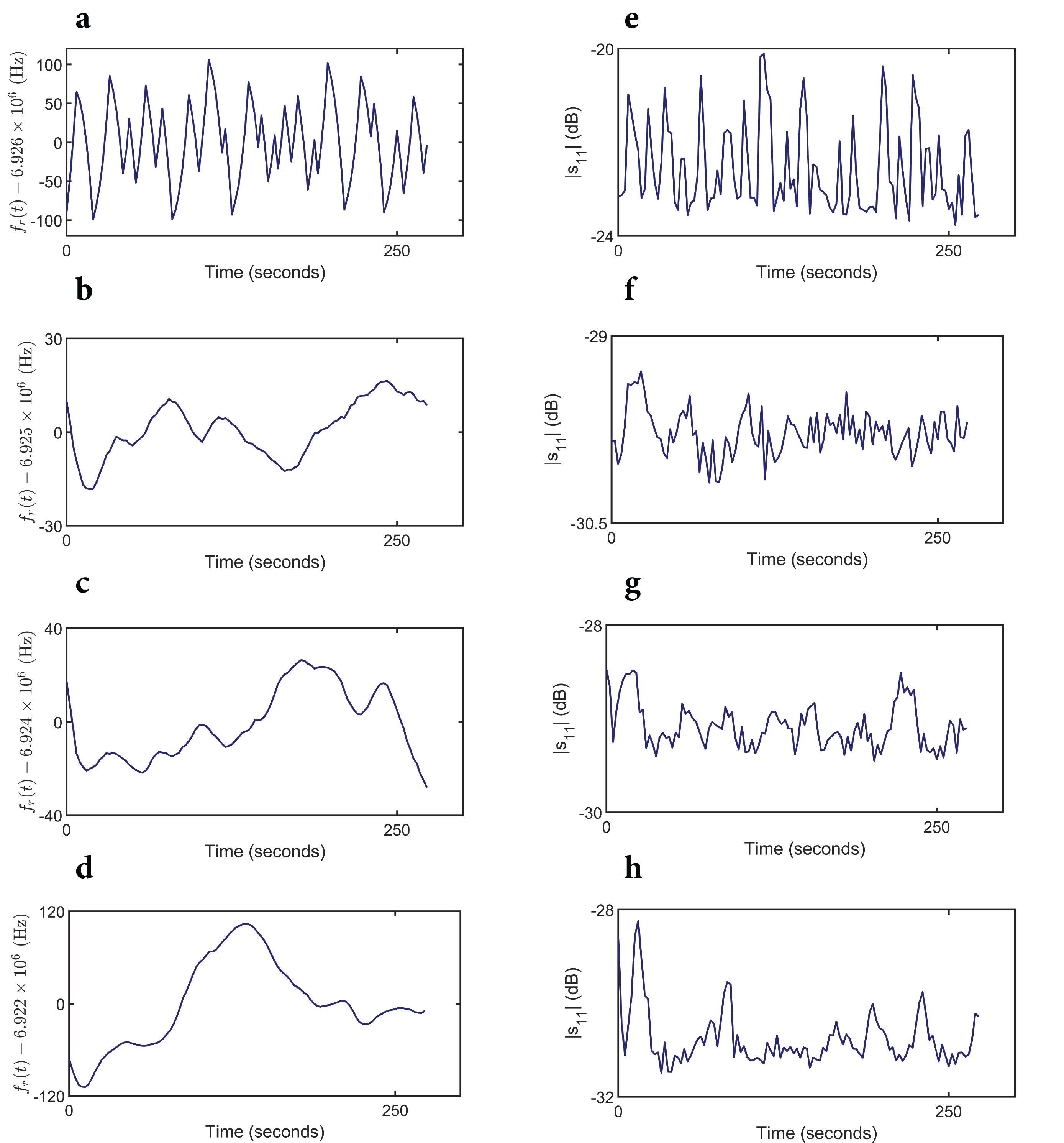}
\caption{Resonant frequency tracking performance. (a) Frequency $f_r(t)$ supplied by the signal generator is shown as a function of time when 0.63~mW of RF power is suppied by the signal generator to the modulator. (b) Frequency $f_r(t)$ supplied by the signal generator is shown as a function of time when 40~mW of RF power is suppied by the signal generator to the modulator. (c) Frequency $f_r(t)$ supplied by the signal generator is shown as a function of time when 90~mW of RF power is suppied by the signal generator to the modulator. (d) Frequency $f_r(t)$ supplied by the signal generator is shown as a function of time when 160~mW of RF power is suppied by the signal generator to the modulator. (e) Time variation of $|s_{11}|$ of the modulator corresponding to (a) is shown. (f) Time variation of $|s_{11}|$ of the modulator corresponding to (b) is shown. (g) Time variation of $|s_{11}|$ of the modulator corresponding to (c) is shown. (h) Time variation of $|s_{11}|$ of the modulator corresponding to (d) is shown.}
\label{fig:sup5}
\end{figure*}

The computed $|s_{11}|$ across time and the variation in $f_r(t)$ for different RF power levels are shown in Fig.~\ref{fig:sup5}. These measurements are each approximately 250~s in duration and correspond only to fine frequency control phases. We observe that the $|s_{11}|$ remains close to its low-power drive value (measured with a VNA with 0~dBm power, see Fig. 1f). The reason the low power measurements in Fig.~\ref{fig:sup5}a and Fig.~\ref{fig:sup5}b have larger variation compared to the other plots is due to the low SNR (20~dB coupling loss from the coupler, low power sent from the signal generator to the modulator, and low reflected power from the modulator). The measurements show that the frequency variation across time is within 1 in 10,0000, with $|s_{11}|$ varying less than 4~dB. These results show that the fine frequency control adequately tracks the electrical resonance of the modulator.

\end{document}